\documentclass[sigconf]{acmart}
\usepackage{microtype}
\usepackage{hyperref}
\usepackage{url}
\usepackage{booktabs}
\usepackage{multirow}
\usepackage{graphicx}
\usepackage{subcaption}
\usepackage{enumitem}
\usepackage{amsmath,amsthm,amsfonts}
\usepackage{xcolor}
\usepackage{makecell}
\usepackage{colortbl}
\definecolor{darkblue}{rgb}{0, 0, 0.5}

\hypersetup{colorlinks=true, citecolor=darkblue, linkcolor=darkblue, urlcolor=darkblue}
\AtBeginDocument{%
  }

\copyrightyear{2026}
\acmYear{2026}
\setcopyright{cc}
\setcctype{by}
\acmConference[SIGIR '26]{Proceedings of the 49th International ACM SIGIR Conference on Research and Development in Information Retrieval}{July 20--24, 2026}{Melbourne, VIC, Australia}
\acmBooktitle{Proceedings of the 49th International ACM SIGIR Conference on Research and Development in Information Retrieval (SIGIR '26), July 20--24, 2026, Melbourne, VIC, Australia}
\acmDOI{10.1145/3805712.3808437}
\acmISBN{979-8-4007-2599-9/2026/07}

\begin{document}

\title{GenRec: A Preference-Oriented Generative Framework for Large-Scale Recommendation}


\author{Yanyan Zou}
\authornote{Core contributors.}
\authornote{Project leader.} 
\affiliation{
  \institution{JD.com}
  \country{Beijing, China}
}
\email{zoe.yyzou@gmail.com}

\author{Junbo Qi}
\authornotemark[1]
\affiliation{%
  \institution{Waseda University}
  \state{Tokyo}
  \country{Japan}
}
\email{junboqi@toki.waseda.jp}

\author{Lunsong Huang}
\authornotemark[1]
\affiliation{%
  \institution{JD.com}
  \country{Beijing, China}
}
\email{huanglunsong1@jd.com}

\author{Yu Li}
\authornotemark[1]
\affiliation{
  \institution{JD.com}
  \country{Beijing, China}
}
\email{liyu.liz@jd.com}

\author{Kewei Xu}
\authornotemark[1]
\affiliation{
  \institution{JD.com}
  \country{Beijing, China}
}
\email{xukewei3@jd.com}

\author{Jiahao Gao}
\authornotemark[1]
\affiliation{
  \institution{JD.com}
  \country{Beijing, China}
}
\email{gaojiahao.20@jd.com}

\author{Binglei Zhao}
\affiliation{
  \institution{JD.com}
  \country{Beijing, China}
}
\email{zhaobinglei1@jd.com}

\author{Xuanhua Yang}
\affiliation{
  \institution{JD.com}
  \country{Beijing, China}
}
\email{yangxuanhua1@jd.com}

\author{Sulong Xu}
\affiliation{
  \institution{JD.com}
  \country{Beijing, China}
}
\email{xusulong@jd.com}

\author{Shengjie Li}
\authornote{Corresponding author.} 
\affiliation{%
  \institution{JD.com}
  \country{Beijing, China}
}
\email{lishengjie1@jd.com}

\renewcommand{\shortauthors}{Yanyan Zou et al.}



\begin{abstract}
Generative Retrieval (GR) offers a promising paradigm for recommendation through next-token prediction (NTP). However, scaling it to large-scale industrial systems introduces three challenges: (i) within
a single request, the identical model inputs may produce inconsistent outputs due to the pagination request mechanism; 
(ii) the prohibitive cost of encoding long user behavior sequences with multi-token item representations based on semantic IDs, and (iii) aligning the generative policy with nuanced user preference signals.
We present \textbf{GenRec}, a preference-oriented generative framework deployed on the JD App~\footnote{\url{https://www.jd.com}} that addresses above challenges within a single decoder-only architecture.
For training objective, we propose \textbf{Page-wise NTP} task, which supervises over an entire interaction page rather than each interacted item individually, providing denser gradient signal and resolving the one-to-many ambiguity of point-wise training.
On the \emph{prefilling} side, an asymmetric linear \textbf{Token Merger} compresses multi-token Semantic IDs in the prompt while preserving full-resolution decoding, reducing input length by ${\sim} \textbf{2} \times$ with negligible accuracy loss.
To further align outputs with user satisfaction, we introduce \textbf{GRPO-SR}, a reinforcement learning method that pairs Group Relative Policy Optimization with NLL regularization for training stability, and employs \textbf{Hybrid Rewards} combining a dense reward model with a relevance gate to mitigate reward hacking.
In month-long online A/B tests serving production traffic, GenRec achieves \textbf{9.5\%} improvement in click count and \textbf{8.7\%} in transaction count over the existing pipeline.
\end{abstract}
\maketitle

\section{Introduction}
Modern recommender systems typically adopt a retrieve-and-rank architecture~\cite{8529185,10494051}.
Recent progresses~\cite{tigergoogle,yang2025sparse,li2025matching} have shown the effectiveness of generative retrieval paradigm.
By reformulating retrieval task as a conditional sequence generation problem, this approach directly generates target items from the entire corpus.
Nevertheless, we observe that deploying such a method in large-scale industrial recommender systems remains challenging.
First, to handle high-volume traffic and ensure user experience, industrial systems typically employ the pagination request mechanism.
Within each paginated request, a user might exhibit multiple positive interactions (e.g., click, transaction), leading to identical model inputs yet multiple valid output in generative task following the vanilla NTP paradigm. 
Second, the long user historical behavior sequences can lead to substantial computational overhead and increased online inference latency.
Third, naively aligning the generative model with personalized objectives could result in reward hacking
and performance degradation.
To address above challenges, we propose \textbf{GenRec}, a generative retrieval-based recommendation framework, unifying user intent understanding and item retrieval within a single decoder-only architecture. It integrates SID-based representations and preference alignment via \textbf{G}roup \textbf{R}elative \textbf{P}olicy \textbf{O}ptimization with \textbf{S}upervised \textbf{R}egularization (GRPO-SR), enabling robust optimization under large-scale real-world users' feedback.

The main contributions are summarized as follows:
\begin{itemize}
    \item We introduce a \textbf{Page-Wise NTP} supervised fine-tuning (SFT) strategy to capture holistic user interaction patterns and resolve the one-to-many ambiguity of vanilla NTP.
    \item We propose an asymmetric representation architecture utilizing a linear \textbf{Token Merger} in \textit{prefilling} side. This mechanism compresses the prompt embedding sequence to efficiently model long user behaviors, while \textit{decoding} side are unmerged SIDs to ensure fine-grained item retrieval.
    \item We develop a reinforcement learning (RL) method  \textbf{GRPO-SR} with \textbf{Hybrid Rewards} for preference alignment, 
    which combines a dense reward model with a gating mechanism that \textbf{suppresses Reward Hacking}. 
    We further introduce Negative Log-Likelihood (NLL) regularization to stabilize training 
    and preserve real-world user behavior patterns.
    \item We empirically demonstrate scaling laws in generative recommendation and validate the framework through large-scale deployment in the JD App, achieving \textbf{9.5\%} click and \textbf{8.7\%} transaction improvements. 
\end{itemize}

\section{Methodology 
}

\subsection{Preliminary 
}


\begin{figure*}[t]
    \centering
    \includegraphics[width=0.85\textwidth]{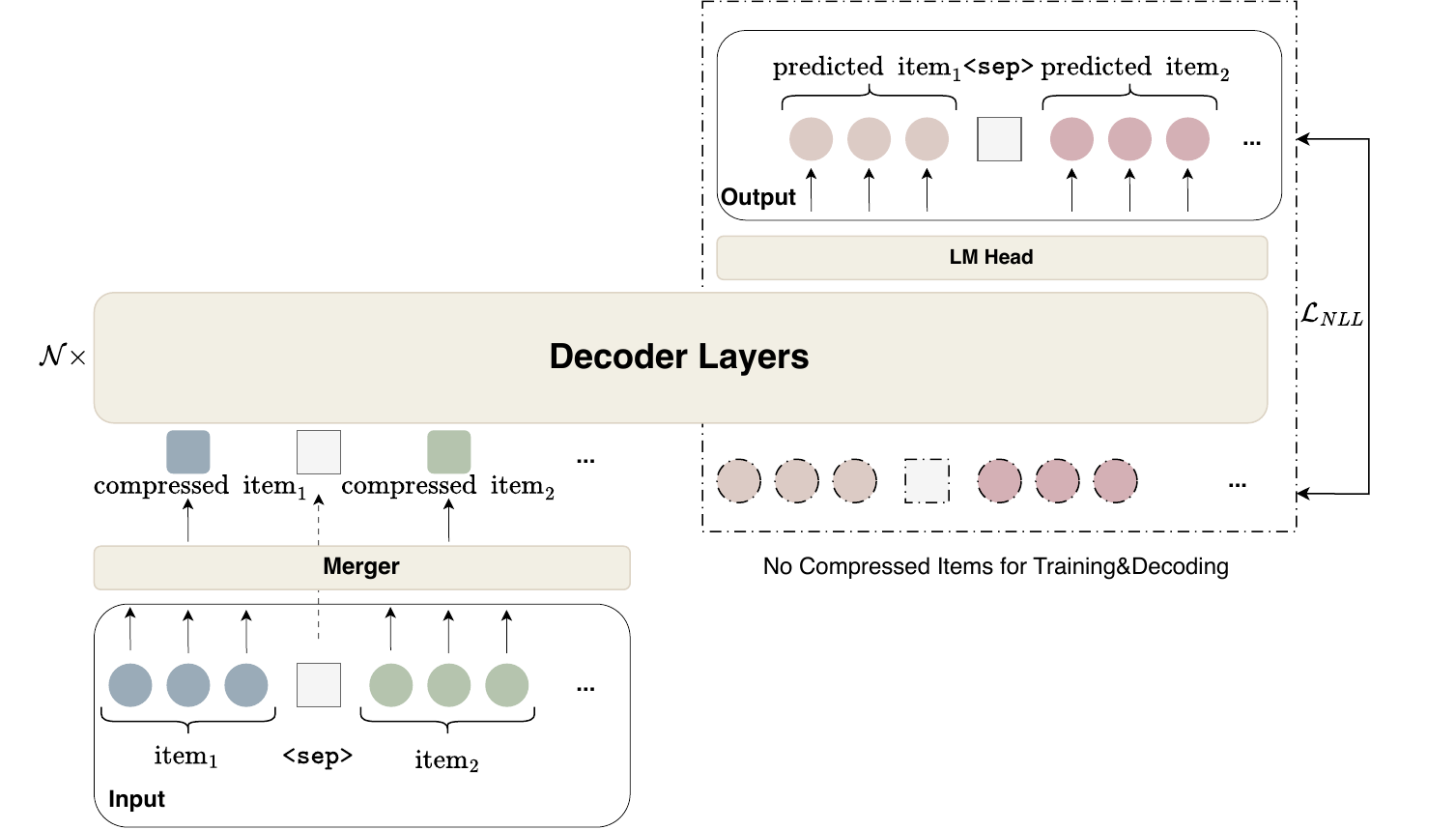} 
    \caption{Model architecture of GenRec. High-dimensional items are quantized into Semantic IDs. To enhance efficiency, an \textbf{Linear token Merger} projects the concatenated embeddings of an item's SIDs into a unified latent vector on the prefilling side. Other tokens (e.g. \texttt{<sep>}) remain uncompressed.}
    \label{fig:teaser} 
\end{figure*}

We formulate the retrieval task as a unified conditional sequence generation problem. Given the  $\mathcal{H}=\{v_1, \ldots, v_n\}$, which denotes the user's historical behavior sequence arranged in chronological order (e.g. $v_n$ represents the most recent interaction), the model predicts the subsequent item $\mathcal{Y}$, representing a potential user interest.

To be specific, the method operates over discrete \emph{Semantic Identifiers} (SIDs)~\cite{tigergoogle}.
A multi-modal model (i.e., Qwen2.5-VL~\cite{bai2025qwen2}) is employed to jointly encode both visual appearance and textual descriptions of each item into a continuous representation to capture more comprehensive item information.
Following existing practices~\cite{onerectechreport}, we fine-tune the embedding model using domain-specific collaborative pairs to ensure the learned embeddings capture recommendation-oriented semantics.
Then, RQ K-means is utilized to discretize the refined embeddings.
This process iteratively clusters the residual vectors, mapping each item $v_i$ to a hierarchical tuple of cluster indices:
\begin{equation}
\label{eq:sid}
\mathrm{SID}(v_i) = \{s_i^1, s_i^2, s_i^3\}.
\end{equation}

\subsection{User-Centric Page-Wise NTP SFT}

Previous generative recommendation methods~\cite{tigergoogle, onerectechreport} train and infer under the same point-wise protocol: predicting a single next item given a user history.
Due to the pagination request mechanism of large-scale industrial recommender systems, we argue that this creates a fundamental \emph{label ambiguity}: when the same history $\mathcal{H}$ is paired with $K$ distinct positive items $\{v^{(k)}\}_{k=1}^{K}$, the model must maximize $\sum_k \log P_\theta(v^{(k)}\mid\mathcal{H})$ over an identical prefix, effectively fitting a uniform mixture over all valid continuations.
This flattened distribution both inflates gradient variance and dilutes per-item probability mass, degrading top-$K$ precision.
The root causes a \emph{cardinality mismatch}: a single session naturally yields multiple engagement signals, yet vanilla point-wise NTP collapses them into isolated input--label pairs, discarding intra-session structure.

To resolve this, we decouple the \emph{training} and \emph{inference} formulations. Following Eq~\ref{eq:sid}, we formulate the input as a composite prompt:
\begin{equation}
S_u= [\mathrm{SID}(v): v \in \mathcal{H}]_{\succ},
\end{equation}

\paragraph{Page-wise Supervision.} 
We further design a page-wise next token prediction (i.e., PW-NTP) strategy for SFT. The target sequence is a \emph{page-wise} list of items the user interacted within the current page, like ordered items $\mathcal{O}$, clicked items $\mathcal{C}$ and exposed items $\mathcal{E}$, ordered by interaction intensity:
%
\begin{equation}
    \label{eq:y_page}
    Y_{\mathrm{page}} = [\mathrm{SID}(v) : v \in \mathcal{O} \cup \mathcal{C} \cup \mathcal{E}]_{\succ}
\end{equation}

The training objective is the standard autoregressive SFT loss over the full response sequence $Y_{\mathrm{page}}$:
\begin{equation}
\label{eq:u2i_nllloss}
\mathcal{L}_{\mathrm{SFT}} = -\sum_{t=1}^{|Y_{\mathrm{page}}|} \log P_\theta(y_t \mid S_u, y_{<t}).
\end{equation}
By supervising over the entire page rather than a single item, each forward pass provides a denser learning signal and resolves the one-to-many ambiguity inherent in point-wise training.

\paragraph{Point-wise Beam Search.}
At serving time, the model generates \emph{beam-width} items per query via beam search, following the standard point-wise protocol. This asymmetry is by design: list-wise training provides richer supervision of model gradients, while point-wise inference maintains compatibility with the production \emph{beam search} pipeline for online serving requirements.

\paragraph{Decoder-only Architecture with Token Merger.} 
To better leverage and reuse the inference optimization techniques developed in the Large Language Model community, we directly adopt a decoder-only transformer architecture, as depicted in Figure~\ref{fig:teaser}.
However, multi-token SIDs triple the input sequence length of the item part, posing severe latency challenges. We address this via a linear \textit{Token Merger}. Since the SID triplet $\{s_i^1, s_i^2, s_i^3\}$ is derived from a single item $v_i$, we concatenate and project their embeddings into a unified vector $\mathbf{h}_{v_i}$ via a linear layer in the prompt part:
\begin{equation}
    \mathbf{h}_{v_i} = \text{Linear}(\text{Concat}(\mathbf{e}(s_i^1), \mathbf{e}(s_i^2), \mathbf{e}(s_i^3))).
\end{equation}
This design compresses item SIDs into latent vectors, reducing prompt length by $\sim2\times$ to accommodate long user sequences within strict inference budgets. Special tokens (e.g., \texttt{<sep>}) are kept unmerged to serve as explicit indicators of structural separation. Crucially, this optimization is confined to the prefilling phase, while the decoding process and generative objective adhere to the original semantic token sequence. 

\subsection{Preference Alignment via RL}

While page-wise SFT captures behavioral regularities from historical logs, it lacks explicit optimization for user satisfaction and is inherently brittle against sparse and non-stationary real-world feedback. To address these limitations, we introduce  a RL method GRPO-SR, building on Group Relative Policy Optimization (GRPO) \cite{shao2024DeepSeekMath}. Unlike PW-NTP SFT stage,
the RL stage aligns with the \emph{point-wise beam search} inference protocol: each rollout generates a single item sequence per query, ensuring consistency between RL training and online serving. Our method optimizes \emph{relative} preferences among multiple generated candidates rather than absolute reward values, thereby improving robustness in industrial settings.

\paragraph{Reward Formulation.}


Raw engagement signals (\emph{e.g.}, clicks) are too sparse to provide effective policy gradients. We instead employ a SIM-based model~\cite{pi2020search} to estimate a continuous preference score $r^{\mathrm{pref}}_i \in [0,1]$ for each rollout candidate $o_i$. A key failure mode is \emph{reward hacking}: the policy produces syntactically valid SID combinations that receive non-trivial $r^{\mathrm{pref}}$ yet are semantically irrelevant. We suppress this with a  gate mechanism $\mathcal{G}_i = \mathbb{I}(s_i > \tau)$ where $\tau$ is a small constant, yielding the hybrid reward:
\begin{equation}
    r_i = \mathcal{G}_i \cdot r^{\mathrm{pref}}_i.
    \label{eq:reward}
\end{equation}

The dense preference model, while smoother than binary labels, may still under-estimate rewards for items the user actually engaged with. Let $\mathcal{D}^{+} = \mathcal{O} \cup \mathcal{C}$ denote the set of ordered and clicked items from the interaction page. We calibrate the reward within each rollout group by anchoring positive items to the group maximum:
\begin{equation}
    \tilde{r}_i = \bigl[1 - \mathbb{I}(o_i \in \mathcal{D}^+)\bigr] \cdot r_i \;+\; \mathbb{I}(o_i \in \mathcal{D}^+) \cdot r_{\max},
    \label{eq:calibration}
\end{equation}
where 
$r_{max}$
is the highest $r_{i}$ in the group. This guarantees that the predicted items that hit real-world user' positive behaviors always receive top-tier rewards, preventing the reward model's estimation bias from down-weighting genuinely preferred items.

\paragraph{GRPO-SR Objective.}
We propose a composite objective that harmonizes group-relative policy optimization with supervised stability. The loss function is defined as:
\begin{equation}
\begin{split}
&\mathcal{L}_{\mathrm{GRPO-SR}}(\theta) = \\
&- \mathbb{E}_{S_u \sim T,\, \{o_i\}_{i=1}^G \sim \pi_{\theta}(\cdot|S_u)} \Bigg[ \frac{1}{G}\sum_{i=1}^G \frac{1}{|o_i|} \sum_{t=1}^{|o_i|} \frac{\pi_\theta(o_{i,t} \mid S_u, o_{i,<t})}{\mathrm{sg}\big(\pi_\theta(o_{i,t} \mid S_u, o_{i,<t})\big)} \hat{A}_{i,t} \Bigg]\\
& - \alpha \cdot \mathbb{E}_{v \sim \mathcal{D}^+} \left[ \sum_{t=1}^{|v|} \log \pi_{\theta}(v_t \mid S_u, v_{<t}) \right]
\end{split}
\label{eq:GRPO-obj}
\end{equation}
where $T$ is training set, and $\hat{A}_{i,t}$ is the advantage derived from group-relative rewards. 
The first term leverages 
a importance sampling $\pi_\theta / \mathrm{sg}(\pi_\theta)$ to enable stable, 
one-step policy updates. 
The second term, weighted by $\alpha$, imposes a negative log-likelihood constraint over positive trajectories $\mathcal{D}^+$. Distinct from standard KL-divergence penalties, this NLL regularizer explicitly anchors the policy to the real-world users' behavior,
mitigating over-optimization towards.

\section{Evaluation}

\subsection{Experimental Setup}
The training and testing datasets are collected from a large-scale recommender platform from the JD.com, covering around 560 million 
user interaction sequences over a one-month period.
We take the data from the last day for testing and the remaining for training.
For SFT, we consider three metrics for evaluation: HitRate (HR@K)~\cite{deshpande2004item}, NDCG (N@K)~\cite{jarvelin2002cumulated} and Hallucination Rate (HaR, the percentage of invalid SIDs of the generated results.). For the RL experiments study, we utilize the highest $r^{\mathrm{SIM}}$ score of the generated K items as Reward Metrics (R@K). 
We consider both traditional and generative methods as our baselines, including BERT4Rec~\cite{sun2019bert4rec}, SASRec~\cite{kang2018self}, TIGER~\cite{tigergoogle}, and LC-Rec~\cite{zheng2024adaptinglargelanguagemodels}. It is worthy noting that both TIGER and LC-Rec are trained via vanilla point-wise NTP task. 
The Qwen2.5~\cite{qwen2025qwen25technicalreport} decoder-only architecture is adopted as our backbone. 
To be specific, our model is further trained on the Qwen2.5 variants of 1.5B, 3B and 7B.
To train the generative models, we conducted distributed training across 8 NVIDIA H100 GPUs.
The AdamW \cite{kingma2014adam} is employed as the optimizer with a linear warm-up phase over the first 1\% of training steps, followed by cosine learning rate decay.


\subsection{Evaluation Results}
\begin{table}[t!]
\centering
\small
\begin{tabular}{lcccccc}
\toprule
Model & HR@1 & HR@10 & N@10 & HR@50 & N@50 & HaR $\downarrow$ \\
\midrule
\rowcolor{gray!25}
\multicolumn{7}{l}{\textit{Traditional Methods}} \\
BERT4Rec     & 0.0315 & 0.0968 & 0.0412 & 0.1832 & 0.0689 & - \\
SASRec       & 0.0383 & 0.1048 & 0.0492 & 0.1976 & 0.0776 & - \\
\rowcolor{gray!25}
\multicolumn{7}{l}{\textit{Generative Methods}} \\
TIGER        & 0.0518 & 0.1660 & 0.0803 & 0.3556 & 0.1409 & 15.46\% \\
LC-Rec       & 0.0947 & 0.3669 & 0.2146 & 0.6226 & 0.2717 & 7.80\% \\
\rowcolor{gray!25}
\multicolumn{7}{l}{\textit{Ours}} \\
GenRec       & \underline{0.1189} & \underline{0.4456} & \underline{0.2635} & \underline{0.7192} & \underline{0.3247} & \underline{4.96\%} \\
\hspace{1mm} w/o TM & \textbf{0.1193} & \textbf{0.4467} & \textbf{0.2653} & \textbf{0.7201} & \textbf{0.3276} & \textbf{4.89\%} \\
\bottomrule
\end{tabular}
\caption{Next-item vs. next-sequence prediction performance. ``TM'' denotes Token Merger. Best results are in \textbf{bold}, and the second best are \underline{underlined}. }
\label{tab:nextitem_nextseq}
\end{table}

\subsubsection{Effectiveness of the Model Structure and SFT Framework.}
\label{sec:main_performance}
We take the Qwen2.5 3B as our main backbone.
To make fair comparison, we reproduce the LC-Rec using the same Qwen2.5 variant. Recall that LC-Rec is trained following vanilla next-token prediction task.
Table~\ref{tab:nextitem_nextseq} illustrates the offline results of various methods on the large-scale industrial dataset.
Our method (denoted as ``GenRec") achieve better performance on HR and N yet lower HaR compared to both traditional and generative methods.
When comparing the variant of our method with the full token as the input of the decoder module (i.e., removing the token merger module, denoted as ``w/o TM"), both the performance and the valid generation rate remain comparable to the one with token merger. 
This demonstrates the effectiveness of the simple token merger where the decisive information of input are preserved while the input token length of the decoder module is reduced by half.

\begin{figure*}[!t]
\centering
\begin{subfigure}[b]{0.98\columnwidth}
\includegraphics[width=\textwidth]{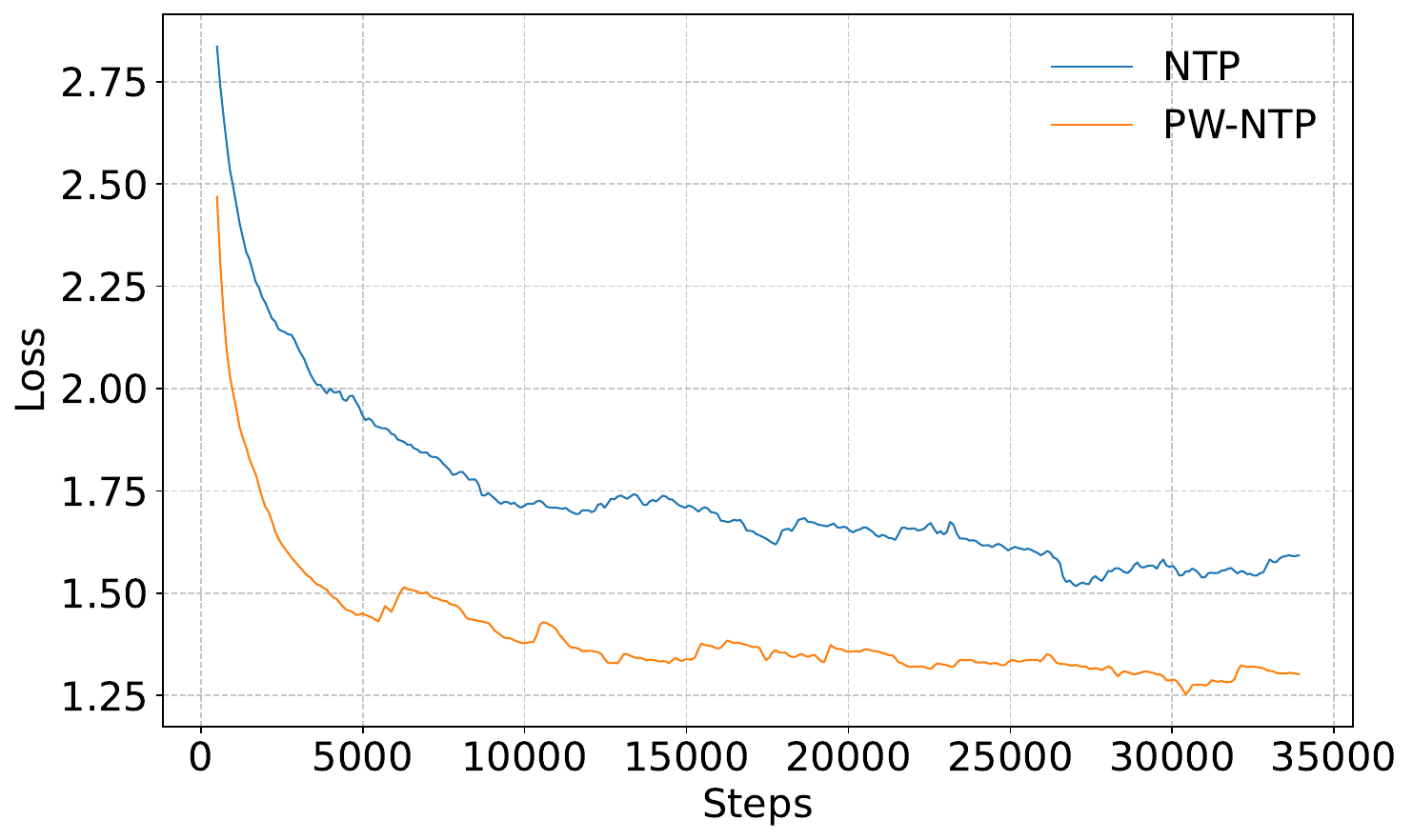}
\caption{Task formulation}
\label{fig:task_formulation}
\end{subfigure}
\hfill
\begin{subfigure}[b]{0.98\columnwidth}
\includegraphics[width=\textwidth]{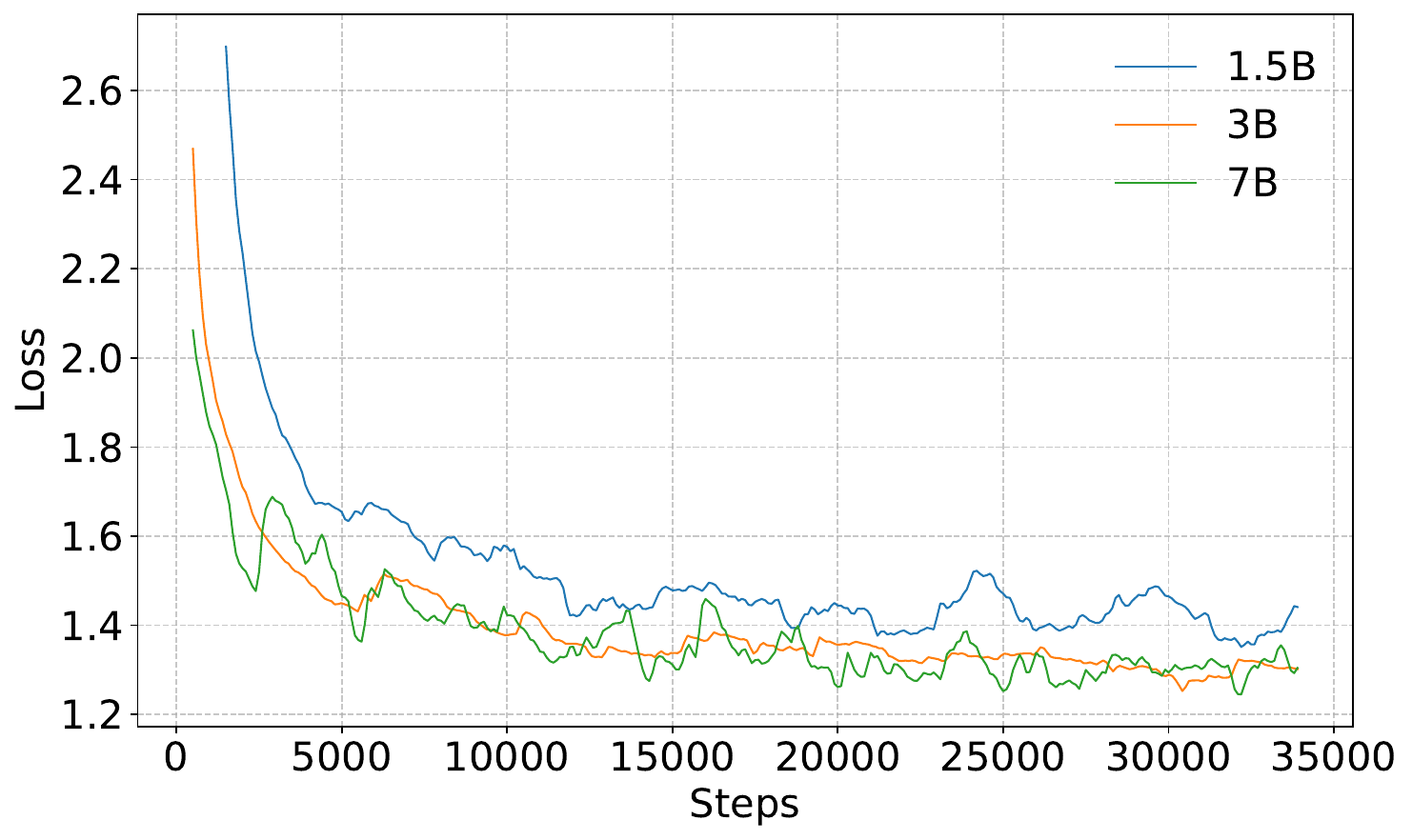}
\caption{Model scaling}
\label{fig:para_scaling}
\end{subfigure}
\caption{SFT loss curves. (a) Page-wise NTP converges faster than NTP. (b) Larger models achieve lower loss with diminishing returns beyond 3B.}
\label{fig:pretraining_dynamics}
\end{figure*}

Furthermore, to investigate the effectiveness of our proposed PW-NTP task, we compare LC-Rec (trained with vanilla NTP), which is equal to the variant of {GenRec} without token merger module as well as trained via vanilla NTP.
As illustrated in Table~\ref{tab:nextitem_nextseq} and Figure~\ref{fig:task_formulation}, GenRec consistently outperforms LC-Rec across all metrics, as well as achieves a better converged loss, demonstrating the effectiveness of the proposed PW-NTP supervision. 

We attribute such significant improvement to two factors:
(1) vanilla NTP creates a one-to-many ambiguity where identical input contexts correspond to multiple valid labels, increasing optimization difficulty and gradient variance; (2) PW-NTP aggregates supervision signals across sequential targets, providing denser learning signal per forward pass and accelerating convergence.
Notably, PW-NTP also reduces the hallucination rate by over 50\%, suggesting that joint prediction encourages more coherent item generation.

\begin{table}[!t]
\centering
\small
\renewcommand{\arraystretch}{1.2}
\begin{tabular}{ccccccc}
\toprule
Model Size & HR@1 & HR@10 & N@10 & HR@50 & N@50 & HaR.$\downarrow$ \\
\midrule
1.5B & 0.1077 & 0.4103 & 0.2484 & 0.6527 & 0.1885 & 5.34\% \\
3B   & \underline{0.1189} & \underline{0.4456} & \underline{0.2635} & \underline{0.7192} & \underline{0.3247} & \textbf{4.96\%} \\
7B   & \textbf{0.1221} & \textbf{0.4483} & \textbf{0.2649} & \textbf{0.7216} & \textbf{0.3269} & \underline{5.42\%} \\
\bottomrule
\end{tabular}
\caption{Model scaling performance. Best results are in \textbf{bold}, and the second best are \underline{underlined}.}
\label{tab:model_scaling}
\end{table}

\subsubsection{Scaling with Model Size}
We investigate model capacity by training Qwen2.5 variants (1.5B, 3B, and 7B) on identical data. As shown in Figure~\ref{fig:para_scaling} and Table~\ref{tab:model_scaling}, training loss consistently decreases with scale. However, the performance gain from 3B to 7B is marginal compared to the 1.5B-to-3B leap, despite a $\sim$2.3$\times$ parameter increase.
Architectural analysis reveals a distinct structural difference: the 3B model is deeper but narrower (36 layers, 2048 hidden size) compared to the 7B variant (28 layers, 3584 hidden size)~\cite{qwen2025qwen25technicalreport}. This suggests that for generative recommendation, the increased depth in the 3B model may allow for more effective modeling of complex user-item interactions through additional non-linear transformations, partially compensating for the lower representational capacity (width). This aligns with the ``capacity density'' hypothesis~\cite{xiao2025densing}, indicating that optimizing depth over width may yield better efficiency for this domain.



\subsubsection{RL Alignment}
We study how RL helps align model outputs with user preferences. By systematically varying the number of generated rollouts and recording the mean reward among candidates, we use Reward Metric of $r^{\mathrm{SIM}}$ as a metric to evaluate the quality of preference alignment.

%


\begin{table}[h]
\centering
\small
 \resizebox{0.48\textwidth}{!}{
\begin{tabular}{lccccc}
\toprule
\multirow{2}{*}{\textbf{Model}} & \multirow{2}{*}{\textbf{HR@50}} & \multicolumn{3}{c}{\textbf{Reward Metrics}} & \multirow{2}{*}{\textbf{HaR} $\downarrow$} \\
\cmidrule(lr){3-5}
 & \textbf{R@1} & \textbf{R@10} & \textbf{R@50} & \\
\midrule
\rowcolor{gray!25}
\multicolumn{6}{l}{\textit{Baseline}} \\
GenRec (Base SFT model)  &0.7192 & 0.1027 & 0.1519 & 0.1776 & 4.96\% \\

\midrule
\rowcolor{gray!25}
\multicolumn{6}{l}{\textit{Policy Gradient Variants}} \\
GRPO &\underline{0.7248} & \underline{0.1177} & \underline{0.1650} & \underline{0.1861} &6.03\% \\
\textbf{GRPO-SR} &\textbf{0.7438} & \textbf{0.1212} & \textbf{0.1679} & \textbf{0.1892} & 2.68\% \\
\midrule
\rowcolor{gray!25}
\multicolumn{6}{l}{\textit{Reward Variants}} \\
GRPO w/o $\mathcal{G}$ &0.6975 & 0.1045 & 0.1608 & 0.1797 & \textbf{1.75\%} \\
GRPO-SR w/o $\mathcal{G}$ &0.7016 & 0.1067 & 0.1598 & 0.1813 & \underline{1.96\%} \\
\bottomrule
\end{tabular}}
\caption{RL performance. We analyze the impact of different method and the gating mechanism ($\mathcal{G}$). Best results are in \textbf{bold}, and the second best are \underline{underlined}.}
\label{tab:pref_align}
\end{table}

Table~\ref{tab:pref_align} demonstrates that the RL-aligned model consistently surpasses the SFT baseline across all inference budgets.
The largest improvement appears at Reward@1 (+18.01\% relative gain) with full GRPO-SR, indicating that RL effectively reshapes the output distribution toward high-reward candidates. However, removing $\mathcal{G}$ causes substantial drops in HR@50 and HaR despite marginal reward gains, which is a clear sign of reward hacking where the model exploits SIM's requirement for valid SIDs that represent the real items while sacrificing overall quality. 
\subsection{Online A/B Testing}
We deploy both base SFT and RL-Aligned models on JD's homepage feed recommendation platform with 10\% traffic each over one month.
As illustrated in Table~\ref{tab:ab_exp}, both versions achieve significant (two-sided test with $p<0.05$) improvements on click and transaction counts.
To be specific, for the long-tail items, the proposed method leads to a 10\% increase in exposure rate, a 16\% increase in click count, and a 13\% increase in transaction count.
GenRec with GRPO-SR alignment is now fully deployed in production.

\begin{table}[!t]
\centering
\setlength{\tabcolsep}{4pt}
\renewcommand{\arraystretch}{1.0}
 \resizebox{0.48\textwidth}{!}{
\begin{tabular}{lccc}
\toprule
Setting & Exposure Rate & Click Count& Transaction Count \\
\midrule
GenRec (Base SFT model)   & 48.7\%  & +8.5\%  & +7.3\%  \\
\hspace{1mm}+ GRPO-SR alignment & 57.3\% & \textbf{+9.5\%} & \textbf{+8.7\%}  \\
\bottomrule
\end{tabular}}
\caption{Online A/B results: RL alignment resolves click-conversion misalignment. 
}
\label{tab:ab_exp}
\end{table}

\section{Related Work}
Motivated by Large Language Models (LLMs), recommender systems are shifting from traditional discriminative to generative modeling~\cite{00dai2025onepiecebringingcontextengineering,01hu2025idssemanticsgenerativeframework,02zhang2025gprgenerativepretrainedonemodel,03zhang2025slowthinkingsequentialrecommendation}. Redefining recommendation as sequence-to-sequence generation, studies optimize backbones: HSTU\cite{04zhai2024actionsspeaklouderwords} boosts efficiency via gated linear recurrence instead of attention, MTGR\cite{05Han_2025} builds an industrial framework on it to balance scalability and precision, and OneTrans \cite{06zhang2025onetransunifiedfeatureinteraction} enhances generalization through multi-task learning and knowledge transfer. 
Treating items as tokens \cite{07chen2024enhancingitemtokenizationgenerative} causes large vocabularies and severe cold-start problems. To address this, Semantic IDs (SIDs) via vector quantization are explored: TIGER \cite{tigergoogle} uses RQ-VAE for cold-start transfer, LETTER \cite{09wang2025learnableitemtokenizationgenerative} optimizes codebooks end-to-end, and OneRec \cite{onerectechreport} adopts iterative RQ K-means for efficient hierarchical IDs. Recent work pursues unified representations \cite{11lin2025unifiedsemanticidrepresentation} to integrate semantic tokens’ generalization and atomic IDs’ specificity.



\section{Conclusion}

This paper presents \textbf{GenRec}, a generative retrival model with preference alignment.
To address industrial challenges, we synergize Multimodal Semantic IDs and an Token Merger for efficient representation learning, while employing Page-wise Generative SFT and GRPO-SR to ensure alignment with hierarchical business objectives.
Large-scale deployment confirms that this differentiable paradigm significantly outperforms traditional multi-stage pipelines, validating its potential as a scalable solution for next-generation recommendation.
We will investigate the reasoning ability of such framework for future work.
\begin{acks}
This work is sponsored by Beijing Nova Program (No.20250484857).
\end{acks}

\subsubsection*{\textbf{Presenter Biography}} Yanyan Zou is an applied scientist in Recommendation Platform at JD.com since 2020, launching cutting-edge AI models into practical productions.
Her research interests primarily lie in the areas of large language model and recommendation, with around 20 papers published in top-tier conferences (e.g., ACL, EMNLP, AAAI). She received her B.Engr. degree in 2015 from Xiamen University, China, as well as her Ph.D. degree from
	Singapore University of Technology and Design
	in 2020.

\bibliographystyle{ACM-Reference-Format}
\balance
\bibliography{ref}

@misc{zhou2025onerectechnicalreport,
      title={OneRec Technical Report}, 
      author={Guorui Zhou and Jiaxin Deng and Jinghao Zhang and Kuo Cai and Lejian Ren and Qiang Luo and Qianqian Wang and Qigen Hu and Rui Huang and Shiyao Wang and Weifeng Ding and Wuchao Li and Xinchen Luo and Xingmei Wang and Zexuan Cheng and Zixing Zhang and Bin Zhang and Boxuan Wang and Chaoyi Ma et al.},
      year={2025},
      eprint={2506.13695},
      archivePrefix={arXiv},
      primaryClass={cs.IR},
      url={https://arxiv.org/abs/2506.13695}, 
}

@inproceedings{Zhao_2025_prerank, series={WWW ’25},
   title={A Hybrid Cross-Stage Coordination Pre-ranking Model for Online Recommendation Systems},
   url={http://dx.doi.org/10.1145/3701716.3715208},
   DOI={10.1145/3701716.3715208},
   booktitle={Companion Proceedings of the ACM on Web Conference 2025},
   publisher={ACM},
   author={Zhao, Binglei and Qi, Houying and Xu, Guang and Ma, Mian and Zhao, Xiwei and Mei, Feng and Xu, Sulong and Hu, Jinghe},
   year={2025},
   month=may, pages={621–630},
   collection={WWW ’25} }

@inproceedings{2022rankflow,
author = {Qin, Jiarui and Zhu, Jiachen and Chen, Bo and Liu, Zhirong and Liu, Weiwen and Tang, Ruiming and Zhang, Rui and Yu, Yong and Zhang, Weinan},
title = {RankFlow: Joint Optimization of Multi-Stage Cascade Ranking Systems as Flows},
year = {2022},
isbn = {9781450387323},
publisher = {Association for Computing Machinery},
address = {New York, NY, USA},
url = {https://doi.org/10.1145/3477495.3532050},
doi = {10.1145/3477495.3532050},
booktitle = {Proceedings of the 45th International ACM SIGIR Conference on Research and Development in Information Retrieval},
pages = {814–824},
numpages = {11},
location = {Madrid, Spain},
series = {SIGIR '22}
}

@inproceedings{2023COPR,
author = {Zhao, Zhishan and Gao, Jingyue and Zhang, Yu and Han, Shuguang and Lou, Siyuan and Sheng, Xiang-Rong and Wang, Zhe and Zhu, Han and Jiang, Yuning and Xu, Jian and Zheng, Bo},
title = {COPR: Consistency-Oriented Pre-Ranking for Online Advertising},
year = {2023},
isbn = {9798400701245},
publisher = {Association for Computing Machinery},
address = {New York, NY, USA},
url = {https://doi.org/10.1145/3583780.3615465},
doi = {10.1145/3583780.3615465},
booktitle = {Proceedings of the 32nd ACM International Conference on Information and Knowledge Management},
pages = {4974–4980},
numpages = {7},
location = {Birmingham, United Kingdom},
series = {CIKM '23}
}

@misc{shao2024DeepSeekMath,
      title={DeepSeekMath: Pushing the Limits of Mathematical Reasoning in Open Language Models}, 
      author={Zhihong Shao and Peiyi Wang and Qihao Zhu and Runxin Xu and Junxiao Song and Xiao Bi and Haowei Zhang and Mingchuan Zhang and Y. K. Li and Y. Wu and Daya Guo},
      year={2024},
      eprint={2402.03300},
      archivePrefix={arXiv},
      primaryClass={cs.CL},
      url={https://arxiv.org/abs/2402.03300}, 
}

@misc{2023TIGER,
      title={Recommender Systems with Generative Retrieval}, 
      author={Shashank Rajput and Nikhil Mehta and Anima Singh and Raghunandan H. Keshavan and Trung Vu and Lukasz Heldt and Lichan Hong and Yi Tay and Vinh Q. Tran and Jonah Samost and Maciej Kula and Ed H. Chi and Maheswaran Sathiamoorthy},
      year={2023},
      eprint={2305.05065},
      archivePrefix={arXiv},
      primaryClass={cs.IR},
      url={https://arxiv.org/abs/2305.05065}, 
}

@article{tigergoogle,
  title={Recommender systems with generative retrieval},
  author={Rajput, Shashank and Mehta, Nikhil and Singh, Anima and Hulikal Keshavan, Raghunandan and Vu, Trung and Heldt, Lukasz and Hong, Lichan and Tay, Yi and Tran, Vinh and Samost, Jonah and others},
  journal={Advances in Neural Information Processing Systems},
  volume={36},
  pages={10299--10315},
  year={2023}
}

@article{bai2025qwen2,
  title={Qwen2. 5-vl technical report},
  author={Bai, Shuai and Chen, Keqin and Liu, Xuejing and Wang, Jialin and Ge, Wenbin and Song, Sibo and Dang, Kai and Wang, Peng and Wang, Shijie and Tang, Jun and others},
  journal={arXiv preprint arXiv:2502.13923},
  year={2025}
}

@misc{00dai2025onepiecebringingcontextengineering,
      title={OnePiece: Bringing Context Engineering and Reasoning to Industrial Cascade Ranking System}, 
      author={Sunhao Dai and Jiakai Tang and Jiahua Wu and Kun Wang and Yuxuan Zhu and Bingjun Chen and Bangyang Hong and Yu Zhao and Cong Fu and Kangle Wu and Yabo Ni and Anxiang Zeng and Wenjie Wang and Xu Chen and Jun Xu and See-Kiong Ng},
      year={2025},
      eprint={2509.18091},
      archivePrefix={arXiv},
      primaryClass={cs.IR},
      url={https://arxiv.org/abs/2509.18091}, 
}

@misc{01hu2025idssemanticsgenerativeframework,
      title={From IDs to Semantics: A Generative Framework for Cross-Domain Recommendation with Adaptive Semantic Tokenization}, 
      author={Peiyu Hu and Wayne Lu and Jia Wang},
      year={2025},
      eprint={2511.08006},
      archivePrefix={arXiv},
      primaryClass={cs.IR},
      url={https://arxiv.org/abs/2511.08006}, 
}

@misc{02zhang2025gprgenerativepretrainedonemodel,
      title={GPR: Towards a Generative Pre-trained One-Model Paradigm for Large-Scale Advertising Recommendation}, 
      author={Jun Zhang and Yi Li and Yue Liu and Changping Wang and Yuan Wang and Yuling Xiong and Xun Liu and Haiyang Wu and Qian Li and Enming Zhang and Jiawei Sun and Xin Xu and Zishuai Zhang and Ruoran Liu and Suyuan Huang and Zhaoxin Zhang and Zhengkai Guo and Shuojin Yang and Meng-Hao Guo and Huan Yu and Jie Jiang and Shi-Min Hu},
      year={2025},
      eprint={2511.10138},
      archivePrefix={arXiv},
      primaryClass={cs.IR},
      url={https://arxiv.org/abs/2511.10138}, 
}

@misc{03zhang2025slowthinkingsequentialrecommendation,
      title={Slow Thinking for Sequential Recommendation}, 
      author={Junjie Zhang and Beichen Zhang and Wenqi Sun and Hongyu Lu and Wayne Xin Zhao and Yu Chen and Ji-Rong Wen},
      year={2025},
      eprint={2504.09627},
      archivePrefix={arXiv},
      primaryClass={cs.IR},
      url={https://arxiv.org/abs/2504.09627}, 
}

@misc{04zhai2024actionsspeaklouderwords,
      title={Actions Speak Louder than Words: Trillion-Parameter Sequential Transducers for Generative Recommendations}, 
      author={Jiaqi Zhai and Lucy Liao and Xing Liu and Yueming Wang and Rui Li and Xuan Cao and Leon Gao and Zhaojie Gong and Fangda Gu and Michael He and Yinghai Lu and Yu Shi},
      year={2024},
      eprint={2402.17152},
      archivePrefix={arXiv},
      primaryClass={cs.LG},
      url={https://arxiv.org/abs/2402.17152}, 
}

@inproceedings{05Han_2025, series={CIKM ’25},
   title={MTGR: Industrial-Scale Generative Recommendation Framework in Meituan},
   url={http://dx.doi.org/10.1145/3746252.3761565},
   DOI={10.1145/3746252.3761565},
   booktitle={Proceedings of the 34th ACM International Conference on Information and Knowledge Management},
   publisher={ACM},
   author={Han, Ruidong and Yin, Bin and Chen, Shangyu and Jiang, He and Jiang, Fei and Li, Xiang and Ma, Chi and Huang, Mincong and Li, Xiaoguang and Jing, Chunzhen and Han, Yueming and Zhou, MengLei and Yu, Lei and Liu, Chuan and Lin, Wei},
   year={2025},
   month=nov, pages={5731–5738},
   collection={CIKM ’25} 
}

@misc{06zhang2025onetransunifiedfeatureinteraction,
      title={OneTrans: Unified Feature Interaction and Sequence Modeling with One Transformer in Industrial Recommender}, 
      author={Zhaoqi Zhang and Haolei Pei and Jun Guo and Tianyu Wang and Yufei Feng and Hui Sun and Shaowei Liu and Aixin Sun},
      year={2025},
      eprint={2510.26104},
      archivePrefix={arXiv},
      primaryClass={cs.IR},
      url={https://arxiv.org/abs/2510.26104}, 
}

@misc{07chen2024enhancingitemtokenizationgenerative,
      title={Enhancing Item Tokenization for Generative Recommendation through Self-Improvement}, 
      author={Runjin Chen and Mingxuan Ju and Ngoc Bui and Dimosthenis Antypas and Stanley Cai and Xiaopeng Wu and Leonardo Neves and Zhangyang Wang and Neil Shah and Tong Zhao},
      year={2024},
      eprint={2412.17171},
      archivePrefix={arXiv},
      primaryClass={cs.LG},
      url={https://arxiv.org/abs/2412.17171}, 
}

@misc{09wang2025learnableitemtokenizationgenerative,
      title={Learnable Item Tokenization for Generative Recommendation}, 
      author={Wenjie Wang and Honghui Bao and Xinyu Lin and Jizhi Zhang and Yongqi Li and Fuli Feng and See-Kiong Ng and Tat-Seng Chua},
      year={2025},
      eprint={2405.07314},
      archivePrefix={arXiv},
      primaryClass={cs.IR},
      url={https://arxiv.org/abs/2405.07314}, 
}

@misc{onerectechreport,
      title={OneRec Technical Report}, 
      author={Guorui Zhou and Jiaxin Deng and Jinghao Zhang and Kuo Cai and Lejian Ren and Qiang Luo and Qianqian Wang and Qigen Hu and Rui Huang and Shiyao Wang and Weifeng Ding and Wuchao Li and Xinchen Luo and Xingmei Wang and Zexuan Cheng and Zixing Zhang and Bin Zhang and Boxuan Wang and Chaoyi Ma and Chengru Song and Chenhui Wang and Di Wang and Dongxue Meng and Fan Yang and Fangyu Zhang and Feng Jiang and Fuxing Zhang and Gang Wang and Guowang Zhang and Han Li and Hengrui Hu and Hezheng Lin and Hongtao Cheng and Hongyang Cao and Huanjie Wang and Jiaming Huang and Jiapeng Chen and Jiaqiang Liu and Jinghui Jia and Kun Gai and Lantao Hu and Liang Zeng and Liao Yu and Qiang Wang and Qidong Zhou and Shengzhe Wang and Shihui He and Shuang Yang and Shujie Yang and Sui Huang and Tao Wu and Tiantian He and Tingting Gao and Wei Yuan and Xiao Liang and Xiaoxiao Xu and Xugang Liu and Yan Wang and Yi Wang and Yiwu Liu and Yue Song and Yufei Zhang and Yunfan Wu and Yunfeng Zhao and Zhanyu Liu},
      year={2025},
      eprint={2506.13695},
      archivePrefix={arXiv},
      primaryClass={cs.IR},
      url={https://arxiv.org/abs/2506.13695}, 
}

@misc{11lin2025unifiedsemanticidrepresentation,
      title={Unified Semantic and ID Representation Learning for Deep Recommenders}, 
      author={Guanyu Lin and Zhigang Hua and Tao Feng and Shuang Yang and Bo Long and Jiaxuan You},
      year={2025},
      eprint={2502.16474},
      archivePrefix={arXiv},
      primaryClass={cs.IR},
      url={https://arxiv.org/abs/2502.16474}, 
}

@misc{qwen2025qwen25technicalreport,
      title={Qwen2.5 Technical Report}, 
      author={Qwen and : and An Yang and Baosong Yang and Beichen Zhang and Binyuan Hui and Bo Zheng and Bowen Yu and Chengyuan Li and Dayiheng Liu and Fei Huang and Haoran Wei and Huan Lin and Jian Yang and Jianhong Tu and Jianwei Zhang and Jianxin Yang and Jiaxi Yang and Jingren Zhou and Junyang Lin and Kai Dang and Keming Lu and Keqin Bao and Kexin Yang and Le Yu and Mei Li and Mingfeng Xue and Pei Zhang and Qin Zhu and Rui Men and Runji Lin and Tianhao Li and Tianyi Tang and Tingyu Xia and Xingzhang Ren and Xuancheng Ren and Yang Fan and Yang Su and Yichang Zhang and Yu Wan and Yuqiong Liu and Zeyu Cui and Zhenru Zhang and Zihan Qiu},
      year={2025},
      eprint={2412.15115},
      archivePrefix={arXiv},
      primaryClass={cs.CL},
      url={https://arxiv.org/abs/2412.15115}, 
}

@article{kingma2014adam,
  title={Adam: A method for stochastic optimization},
  author={Kingma, Diederik P},
  journal={arXiv preprint arXiv:1412.6980},
  year={2014}
}

@article{deshpande2004item,
  title={Item-based top-n recommendation algorithms},
  author={Deshpande, Mukund and Karypis, George},
  journal={ACM Transactions on Information Systems (TOIS)},
  volume={22},
  number={1},
  pages={143--177},
  year={2004},
  publisher={ACM New York, NY, USA}
}

@article{jarvelin2002cumulated,
  title={Cumulated gain-based evaluation of IR techniques},
  author={J{\"a}rvelin, Kalervo and Kek{\"a}l{\"a}inen, Jaana},
  journal={ACM Transactions on Information Systems (TOIS)},
  volume={20},
  number={4},
  pages={422--446},
  year={2002},
  publisher={ACM New York, NY, USA}
}

@inproceedings{sun2019bert4rec,
  title = {BERT4Rec: Sequential Recommendation with Bidirectional Encoder Representations from Transformers},
  author = {Sun, Fei and Liu, Jun and Wu, Jian and Pei, Changhua and Lin, Xiao and Ou, Wenwu and Jiang, Peng},
  booktitle = {CIKM},
  year = {2019}
}

@inproceedings{kang2018self,
  title={Self-attentive sequential recommendation},
  author={Kang, Wang-Cheng and McAuley, Julian},
  booktitle={2018 IEEE international conference on data mining (ICDM)},
  pages={197--206},
  year={2018},
  organization={IEEE}
}

@inproceedings{zheng2024adaptinglargelanguagemodels,
  title={Adapting large language models by integrating collaborative semantics for recommendation},
  author={Zheng, Bowen and Hou, Yupeng and Lu, Hongyu and Chen, Yu and Zhao, Wayne Xin and Chen, Ming and Wen, Ji-Rong},
  booktitle={2024 IEEE 40th International Conference on Data Engineering (ICDE)},
  pages={1435--1448},
  year={2024},
  organization={IEEE}
}

@article{xiao2025densing,
  title={Densing law of llms},
  author={Xiao, Chaojun and Cai, Jie and Zhao, Weilin and Lin, Biyuan and Zeng, Guoyang and Zhou, Jie and Zheng, Zhi and Han, Xu and Liu, Zhiyuan and Sun, Maosong},
  journal={Nature Machine Intelligence},
  pages={1--11},
  year={2025},
  publisher={Nature Publishing Group UK London}
}

@inproceedings{pi2020search,
  title={Search-based user interest modeling with lifelong sequential behavior data for click-through rate prediction},
  author={Pi, Qi and Zhou, Guorui and Zhang, Yujing and Wang, Zhe and Ren, Lejian and Fan, Ying and Zhu, Xiaoqiang and Gai, Kun},
  booktitle={Proceedings of the 29th ACM International Conference on Information \& Knowledge Management},
  pages={2685--2692},
  year={2020}
}

@ARTICLE{8529185,
  author={Mu, Ruihui},
  journal={IEEE Access}, 
  title={A Survey of Recommender Systems Based on Deep Learning}, 
  year={2018},
  volume={6},
  number={},
  pages={69009-69022},
  doi={10.1109/ACCESS.2018.2880197}}

@ARTICLE{10494051,
  author={Li, Yang and Liu, Kangbo and Satapathy, Ranjan and Wang, Suhang and Cambria, Erik},
  journal={IEEE Computational Intelligence Magazine}, 
  title={Recent Developments in Recommender Systems: A Survey [Review Article]}, 
  year={2024},
  volume={19},
  number={2},
  pages={78-95},
  doi={10.1109/MCI.2024.3363984}}

@article{yang2025sparse,
  title={Sparse meets dense: Unified generative recommendations with cascaded sparse-dense representations},
  author={Yang, Yuhao and Ji, Zhi and Li, Zhaopeng and Li, Yi and Mo, Zhonglin and Ding, Yue and Chen, Kai and Zhang, Zijian and Li, Jie and Li, Shuanglong and others},
  journal={arXiv preprint arXiv:2503.02453},
  year={2025}
}

@article{li2025matching,
  title={From matching to generation: A survey on generative information retrieval},
  author={Li, Xiaoxi and Jin, Jiajie and Zhou, Yujia and Zhang, Yuyao and Zhang, Peitian and Zhu, Yutao and Dou, Zhicheng},
  journal={ACM Transactions on Information Systems},
  volume={43},
  number={3},
  pages={1--62},
  year={2025},
  publisher={ACM New York, NY}
}
\end{document}